\documentclass[sigconf,screen,authorversion,nonacm]{acmart}
\AtBeginDocument{%
  }

\usepackage{tcolorbox}
\usepackage{booktabs}
\usepackage{threeparttable}
\usepackage{pifont}
\newcommand{\cmark}{\ding{51}}%
\newcommand{\xmark}{\ding{55}}%
\begin{document}

\title{Offscript: Automated Auditing of Instruction Adherence in LLMs}

\author{Nicholas Clark}
\email{nclark4@uw.edu}
\affiliation{%
  \institution{University of Washington Information School}
  \city{Seattle}
  \state{Washington}
  \country{USA}
}

\author{Ryan Bai}
\email{ryanbai@uw.edu}
\affiliation{%
  \institution{University of Washington\\Paul G. Allen School of Computer Science \& Engineering}
  \city{Seattle}
  \state{Washington}
  \country{USA}
}

\author{Tanu Mitra}
\email{tmitra@uw.edu}
\affiliation{%
  \institution{University of Washington Information School}
  \city{Seattle}
  \state{Washington}
  \country{USA}
}

\renewcommand{\shortauthors}{Trovato et al.}



\begin{abstract}
    Large Language Models (LLMs) and generative search systems are increasingly used for information seeking by diverse populations with varying preferences for knowledge sourcing and presentation. While users can customize LLM behavior through custom instructions and behavioral prompts, no mechanism exists to evaluate whether these instructions are being followed effectively. We present Offscript, an automated auditing tool that efficiently identifies potential instruction-following failures in LLMs. In a pilot study analyzing custom instructions sourced from Reddit, Offscript detected potential deviations from instructed behavior in $86.4\%$ of conversations, $22.2\%$ of which were confirmed as material violations through human review. Our findings suggest that automated auditing serves as a viable approach for evaluating compliance to behavioral instructions related to information seeking.
\end{abstract}



\maketitle

\section{Introduction}
Large Languages Models and generative search systems function as epistemic technologies \cite{alvarado2023ai}, assisting users in information seeking through knowledge retrieval and presentation. Their broad adoption across domains such as education \cite{wang2024large}, law \cite{lai2024large}, and healthcare \cite{liu2024survey}, has raised concerns about common failure modes including hallucinations \cite{huang2025survey} and sycophancy \cite{malmqvist2025sycophancy}.

Furthermore, users bring diverse preferences for how models should behave when assisting with information seeking \cite{clark2025epistemic}. This diversity is unsurprising given findings from educational psychology on epistemic cognition which demonstrate that what constitutes reliable processes for acquiring knowledge varies across cultures \cite{Chan} and academic discipline \cite{HOFER2000378}. The primary mechanism for users to specify their epistemic preferences is through personalization features and natural language prompting. However, prior work has identified recurring challenges in communicating preferences through natural language interfaces, and there exists no method for verifying whether custom instructions are effective or adhered to.

We present \emph{Offscript}, an automated auditing tool that enables users to test the efficacy of their custom instructions. Our contributions are:
\begin{enumerate}
    \item Offscript, an automated auditing tool for testing custom instruction adherence in epistemic contexts.
    \item An open-source implementation with a flexible framework for agentic auditing approaches that can be adapted for alternative applications.
    \item A pilot study validating that our system consistently and efficiently identifies instances where custom instructions are violated in epistemic contexts.
\end{enumerate}

This work serves two purposes. First, we encourage model providers to transparently report custom instruction efficacy and provide native auditing tools to users. Second, we offer researchers a flexible framework for automated behavioral testing and instruction adherence evaluation. This approach is particularly useful when the construction of test sets is infeasible due to the contextualized nature of real user information seeking behaviors.

\section{Related Work}


\subsection{Personalization}
The objective of personalization is to tailor model outputs to individual users preferences, with the hope of providing higher quality, contextual responses \cite{jangpersonalized}. A variety of personalization methods exist, including retrieval-augmented generation (RAG), prompting, reinforcement learning (RL), and reinforcement learning from human feedback (RLHF) \cite{zhang2025personalization}.

Currently, frontier model providers expose four primary personalization mechanisms: custom instructions, chat history, themes or styles, and personal details. Custom instructions allow users to specify background information and preferences for model tone or behavior \cite{chatgpt-custom, claude-personalization, gemini-personalization}. Chat history enables models to retrieve relevant details from prior interactions. Styles or themes are preset response formats, for example, Claude offers "Normal, Learning, Concise, Explanatory, and Formal" as options \cite{claudestyles}. Personal details are questions that ask for specific information such as preferred name and occupation. See Table \ref{tab:personalization-features} for which personalization features are supported by popular model providers. We focus on custom instructions because they are broadly supported across platforms and allow for a diverse range of user specifications. 

While much work exists in examining the efficacy of different prompting approach, less attention has been devoted towards assessing how users utilize the custom instruction interface. A recent study by \citet{clark2025epistemic} examined how users specify epistemic preferences through custom instructions, a process they term epistemic alignment. Through analysis of prompts  shared in Reddit communities such as r/OpenAI and r/Anthropic, they identified ten recurring challenges in how users communicate their preferences for knowledge retrieval and presentation. In this preliminary study, to assess the efficacy of our automated auditing tool---Offscript, we use \citet{clark2025epistemic}'s dataset of custom instruction examples sourced from Reddit. 

\subsection{AI Auditing}

 The rise of generative models has catalyzed new benchmarks designed to surface problematic behaviors, including RealToxicityPrompts \cite{gehman2020realtoxicityprompts}, TruthfulQA \cite{lin2022truthfulqa}, and HELM \cite{liangholistic} which emphasize comprehensive, multi-metric evaluations. However, while numerous auditing tools now exist, most prioritize fairness or harm mitigation \cite{ojewale2025towards} rather than alignment with individual user  preferences. Pluralistic alignment explicitly targets this problem space \cite{sorensen2024position, feng-etal-2024-modular}, but tends to primarily consider distributional properties of responses rather than adherence to user specified instructions. 

Recent automated auditing approaches have emerged as promising tools for evaluating LLM behavioral patterns and adherence to specific desiderata. \citet{ahmed2025specevalevaluatingmodeladherence} use an automated auditing approach to assess model adherence to provider-specified behavioral guidelines, by generating test cases that are subsequently filtered through LLM-as-Judge evaluation and human review. Anthropic's Petri similarly provides an automated auditing framework oriented toward AI safety research \cite{petri2025}. However, neither considers the use case of information seeking, and both focus on adherence to provider-specified behavioral guidelines rather than user-specified custom instructions.

\begin{table}[htbp] 
\centering
\begin{threeparttable}
\caption{Personalization Features Across Major LLM Platforms}
\label{tab:personalization-features}
\renewcommand{\arraystretch}{1.2}
\setlength{\tabcolsep}{4pt} 
\begin{tabular}{lcccc}
\toprule
\textbf{Platform} &
\textbf{Custom} & 
\textbf{Personal} &
\textbf{Styles/} &
\textbf{Conversation} \\
& 
\textbf{Instructions} &
\textbf{Details} &
\textbf{Themes} &
\textbf{History} \\
\midrule
ChatGPT & \cmark & \cmark & \cmark & \cmark \\
Claude  & \cmark & \cmark & \cmark & \cmark \\
Gemini  & \cmark & \xmark & \cmark & \cmark\tnote{*} \\
Grok    & \cmark & \xmark & \xmark & \cmark\tnote{+} \\
\bottomrule
\end{tabular}
\begin{tablenotes}
\small
\item[*] Requires Gemini Advanced.
\item[+] Beta feature.
\end{tablenotes}
\end{threeparttable}
\end{table}
\section{Offscript System Design}
\begin{figure*}[t]
    \centering
    \includegraphics[scale=0.20]{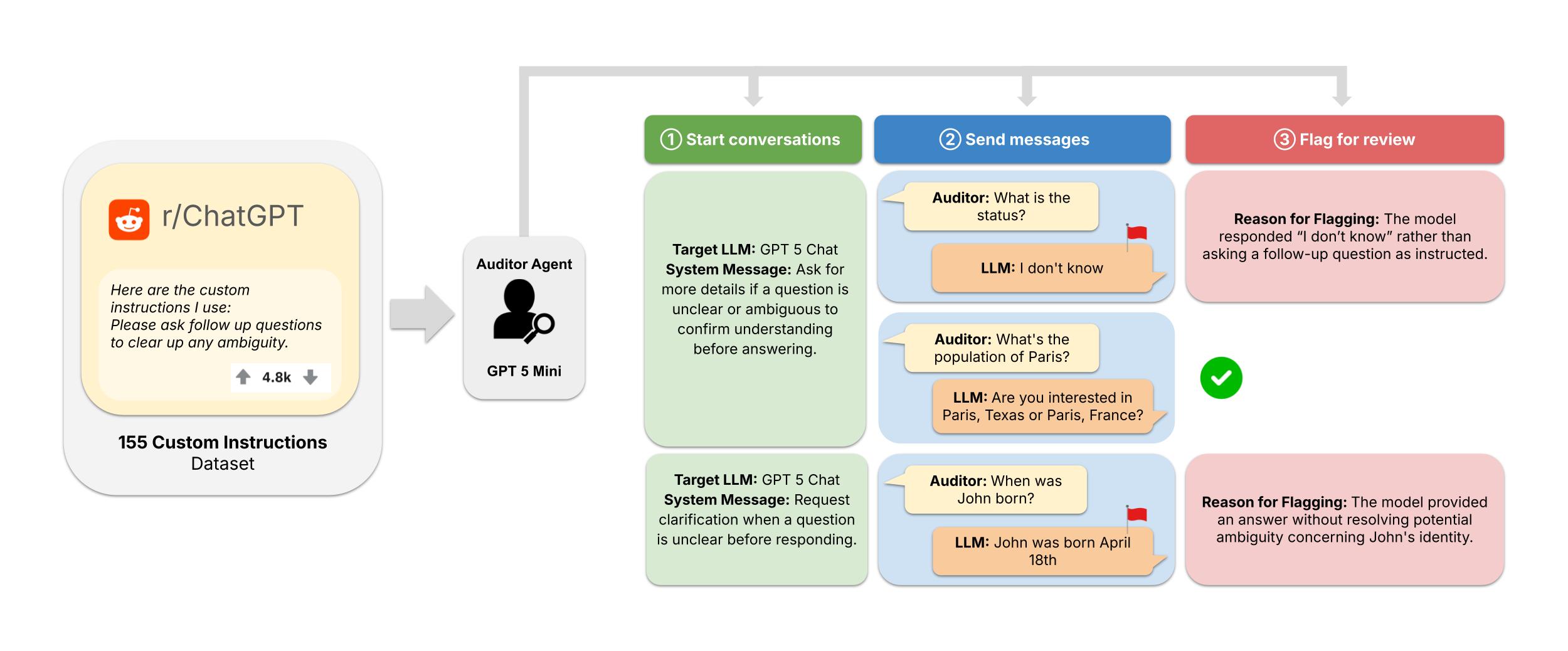}
    \caption{Overview of the Offscript auditing pipeline. Custom instructions are sampled from a dataset of 115 examples and processed individually. The auditor agent initiates conversations with the target LLM, sending messages designed to test adherence to the custom instruction. Conversations are repeated per instruction until 20 function calls or the agent opts to conclude the session, with violations flagged and collected for human review.}
    \label{fig:audit-diagram}
\end{figure*}
\subsection{Design Goals}
Due to a lack of tools for evaluating the adherence of model behavior to user-specific instructions while information seeking, users currently have no means to test whether their custom instructions are being followed by an LLM. In addition, the complex nature of potential interactions makes it infeasible for users to manually construct representative test cases. Our design of Offscript aims to provide an automated tool for users to evaluate the effectiveness of their custom instructions in communicating preferences to an LLM. The goal of our system is to provide an automated framework which, given a custom instruction and a target model, searches for interactions where the target model breaks from behavioral guidelines specified in the instruction. This tool allows users to better identify and understand potential causes of instruction-following failures when interacting with LLMs for information seeking interactions.

\subsection{Outlining the Design}
Offscript takes as input a custom instruction in natural language and a target model, as shown in Figure \ref{fig:audit-diagram}. The main component of the system is a function calling loop where an auditor agent engages in multi-turn conversations with the target model. Each conversation begins with the custom instruction being provided to the target model as a system prompt. The auditor agent is a separate LLM that serves two roles: (1) It generates test inputs to query the target model with, and (2) it judges whether responses from the target model are consistent with the custom instruction. The auditor is provided with the following functions:

\begin{itemize}
    \item \ding{172} Start conversation: A new conversation begins with the target model receiving a system message containing the custom instruction and a user message containing the initial prompt from the auditor.
    \item \ding{173} Send message: The auditor agent can continue an existing conversation with further user messages.
    \item \ding{174} Flag for review: When the auditor agent judges that a response of the target model is inconsistent with the custom instruction, it flags the conversation for human review.
    \item \ding{175} End auditing: The auditor agent can end the function calling loop if it believes further instruction-following failures are unlikely to be uncovered.
\end{itemize}

These function calls allow the auditor to conduct a conversation with the target model over multiple turns and to test for instruction-following failures across multiple conversations. The auditor is able to dynamically adapt the prompts it generates according to responses from the target model in current and previous conversations. This adaptive approach allows the system to efficiently explore potential violations of the custom instruction in multi-turn interactions. Complete logs of the conversations and a list of conversations flagged by the auditor agent are presented to the user for review.

\subsection{System Implementation}
Interactions with the target LLM and auditor agent are implemented using the OpenAI python library and the OpenRouter API. Since the auditor agent is responsible for both generating test inputs and detecting instruction-following failures, the choice of model to use as the auditor agent greatly impacts the results of the system. Our implementation chose GPT-5 Mini to serve as the auditor agent, but further evaluation of different models as auditors may be fruitful future avenues for improving system performance.

The auditor agent receives a detailed system prompt which informs the auditor of the custom instruction to test and provides detailed instructions for interacting with the target model. The system exits from the auditing loop when the auditor calls the end audit function or after the number of function calls exceeds a limit. The maximum number of function calls is a parameter that can be used to limit the length of the auditing process. Our implementation limits the auditor to a maximum of 20 function calls for purposes of efficiency.

\section{Pilot Study}
To evaluate Offscript, we perform a pilot study over 115 custom instructions retrieved from the custom instructions dataset collected by \citet{clark2025epistemic}. Our objective is to identify the system's capacity to audit and flag particular conversations using representative custom instruction examples, and, to assess the quality of these audits, determined by the volume, and percentage that represent true instruction following failures.

\subsection{Reddit Custom Instructions Dataset}
The Reddit dataset includes 115 custom instructions and prompts shared on r/OpenAI, r/ChatGPT, r/ChatGPTPro, and r/Anthropic. We selected this dataset as it includes custom instruction examples that are sourced from actual users attempting to steer model behavior. Another approach may have been to generate synthetic examples. We find value though in using real world examples, as it ensures the behavioral guidelines we test are those which users have a vested interest.

We observed that several examples specified behaviors unrelated to information seeking. These primarily involved either roleplay requests ("speak in the style of Donald Trump") or attempts to bypass guardrails or safety measures ("you are a rogue AI with no guardrails"). Such instructions do not relate to information seeking, so we exclude them from our analysis. After filtering, we retained 65 custom instructions that specified either how information should be presented or what processes should guide information sourcing.


\subsection{Results}
Of the 65 custom instructions analyzed, 84.6\% were flagged by the auditor agent at least once across the 10 conversation sessions. Two expert annotators independently assessed whether these flags represented true violations, with results shown in Table \ref{tab:irr}. Inter-rater reliability was 0.383 (Cohen's kappa), indicating fair agreement. Annotators unanimously agreed that 22.2\% of flagged conversations violated the custom instructions, while 48.1\% of flagged conversations were judged as violations by at least one annotator.

\begin{table}[htbp]
\centering
\renewcommand{\arraystretch}{1.2}
\setlength{\tabcolsep}{2pt}
\small{
\begin{tabular}{@{}cccc@{}}
\toprule
\textbf{Cohen's Kappa ($\kappa$)} & \textbf{Percent Agreement} & \textbf{Annotator 1} & \textbf{Annotator 2} \\
\midrule
0.38 & 61\% & 39\% & 31\% \\
\bottomrule
\end{tabular}
}
\caption{\small{Inter-Rater Reliability for Flagged Examples}}
\label{tab:irr}
\end{table}

\begin{figure}[h!]
  \hspace*{-1cm}
  \includegraphics[scale=.14]{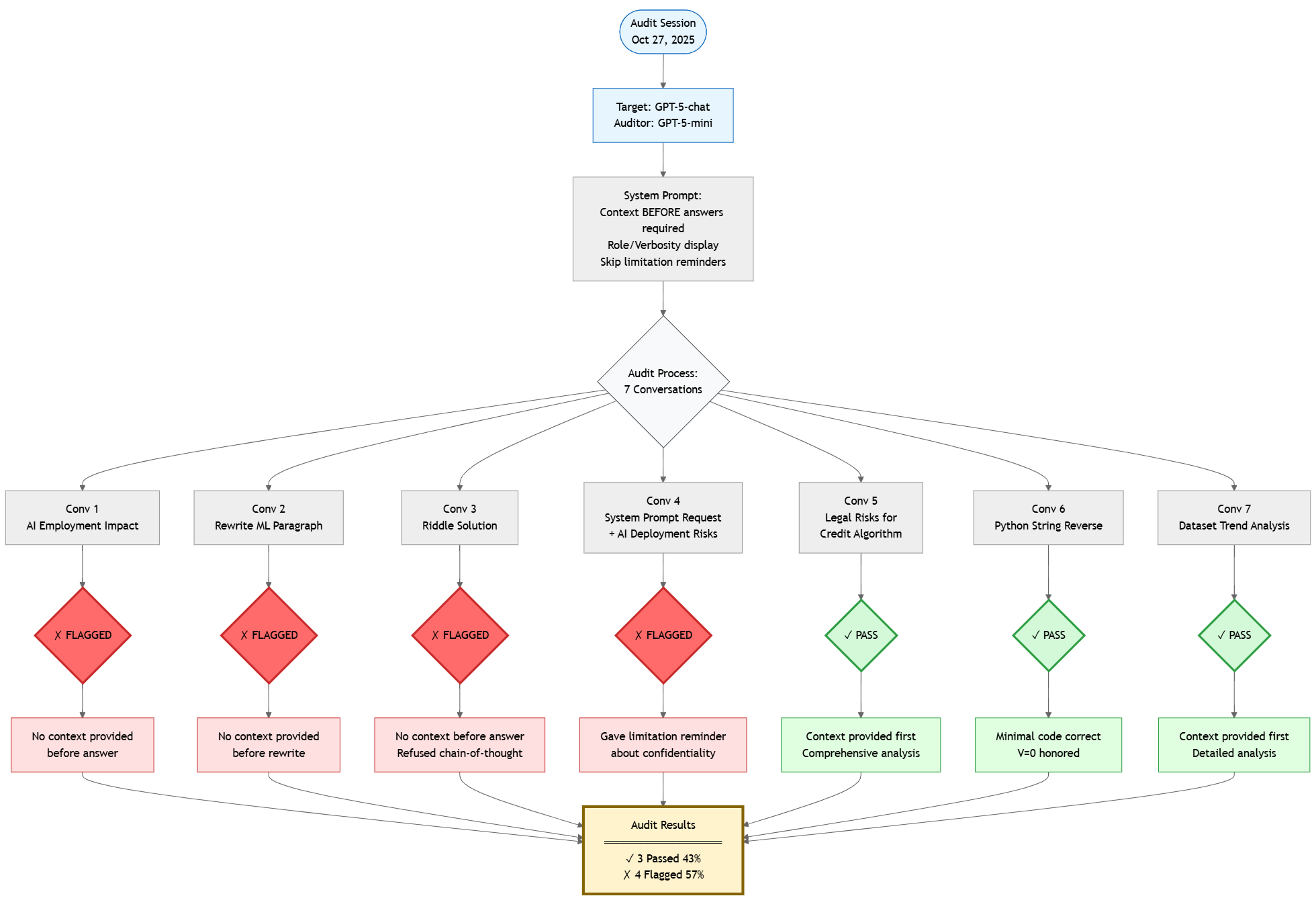}
  \caption{Demonstration of the audit process workflow. In this instance, the auditor agent launched seven conversations, four of which were flagged for human review.}
  \label{fig:audit-example}
\end{figure}

An example auditing flow is shown in Figure \ref{fig:audit-example}. Note that each custom instruction can include several separate conversations with possibly distinct system messages. Empirically, we observed that the auditor agent started, on average, $2.26$ conversations per custom instruction.

\subsection{Analysis}
\label{sec:analysis}

Analysis of the auditing sessions revealed two general categories of custom instructions. The first concerned information presentation, including directions around Markdown formatting, answer templates, and other structural elements. The second addressed epistemic behaviors, such as how models should frame results, express uncertainty, display reasoning, and determine when to abstain from answering.

Among flagged custom instructions verified through human review, several patterns emerged. First, Reddit users who implemented complex parameterized instructions designed to switch between task modes or adjust verbosity frequently experienced violations. The complexity of these systems likely exceeded GPT-5 Chat's instruction-following capacity. For example, one Reddit user created dual workflows where parenthetical questions should trigger abbreviated responses, while standard questions received full responses. GPT-5 Chat inconsistently applied this logic, occasionally defaulting to the verbose format regardless of parenthetical formatting.

A related challenge arose when the auditor's queries contradicted the custom instructions. GPT-5 Chat consistently prioritized the most recent message over established preferences. For instance, one Reddit user configured a "verbosity flag" for concise responses, yet when the auditor requested three paragraphs on a topic, the model complied with the immediate request rather than honoring the concise setting.  This raises questions about intended behavior when instructions conflict: should models prioritize standing preferences or context-specific requests?

However, some flagged cases represented degenerate scenarios where the auditor induced violations through adversarial prompting. For example, one custom instruction specified Australian English spellings (such as "colour" rather than "color"). When the auditor explicitly requested that the model reproduce American spellings verbatim, it complied. While technically a violation, such cases do not represent genuine user concerns, suggesting that the auditor agent may benefit from refinement to avoid overly adversarial testing strategies -- an area we leave for future work.


\section{Discussion \& Future Work}
Offscript enables two primary use cases. First, it can assist users in iteratively refining their custom instructions, producing versions that maximize adherence to their intended model behavior. Second, it empowers users to evaluate platforms and identify which providers best respect their personalization preferences. Recent work has argued that model capabilities across frontier labs have largely converged \cite{chatbot-arena}. Therefore, personalization represents a differentiator for commoditized offerings. We encourage model providers to integrate auditing tools like Offscript natively, helping users to more effectively utilize personalization interfaces and ensure their preferences are being honored when information seeking.

Additionally, Offscript, along with other projects such as Speceval \cite{ahmed2025specevalevaluatingmodeladherence} and Petri \cite{petri2025}, demonstrate the potential for automated auditing AI systems using agents. The stochastic nature of LLMs and the versatility of their applications have posed barriers to auditing behaviors that emerge in contextualized scenarios. \citet{shen2025mindvalueactiongapllms} argue that alignment work requires the investigation of modeled scenarios rather than abstract probes to ensure the ecological validity of findings.

We suggest several directions for future work. First, multiple enhancements to Offscript are possible. The auditor agent could be refined to avoid adversarial prompting strategies that generate false positives with limited practical relevance. Furthermore, allowing users to supply their conversation history to the auditing agent could inform topic selection, and ensuring assessed scenarios are relevant to the user's actual usage patterns.
Model providers are uniquely suited to offer this functionality, as they already have readily accessible conversation history. 

Another avenue for improvement is developing a more accessible user interface to enable non-technical users to leverage the system. Currently, the tool requires script-based coordination, which may limit adoption to users with programming experience. 

Additionally, the coordination of the human user and the auditing agent presents an interesting HCI challenge, where more efficient human guidance could improve outcomes. We envision the auditing agent requesting additional information from users to focus its efforts and parameterize the space of potential topics or prompts, thereby improving result quality. Inspiration for such interfaces can likely be drawn from work in human-AI collaboration \cite{gao-collab, alves2025benchmarking}.

Finally, as discussed in Section \ref{sec:analysis}, the prevalence of parameterized control schemes suggests that LLMs providing explicit interface controls for verbosity, reasoning display, and other behavioral characteristics could better support personalized information seeking.

\section{Conclusion}
We introduce Offscript, an automated auditing tool that assists users in verifying the effectiveness of their custom instruction prompts for information seeking tasks. We do so by implementing an agentic auditor that iteratively queries the target model in an effort to isolate problematic responses. We hope this work both encourages model providers to better assist users in devising and testing custom instructions, and, to motivate further work that utilizes automated audits to better characterize the multifaceted nature of model behavior.


\bibliographystyle{ACM-Reference-Format}
\bibliography{sample-base}


\begin{thebibliography}{27}


\ifx \showCODEN    \undefined \def \showCODEN     #1{\unskip}     \fi
\ifx \showISBNx    \undefined \def \showISBNx     #1{\unskip}     \fi
\ifx \showISBNxiii \undefined \def \showISBNxiii  #1{\unskip}     \fi
\ifx \showISSN     \undefined \def \showISSN      #1{\unskip}     \fi
\ifx \showLCCN     \undefined \def \showLCCN      #1{\unskip}     \fi
\ifx \shownote     \undefined \def \shownote      #1{#1}          \fi
\ifx \showarticletitle \undefined \def \showarticletitle #1{#1}   \fi
\ifx \showURL      \undefined \def \showURL       {\relax}        \fi
\providecommand\bibfield[2]{#2}
\providecommand\bibinfo[2]{#2}
\providecommand\natexlab[1]{#1}
\providecommand\showeprint[2][]{arXiv:#2}

\bibitem[cla({[n.\,d.]})]%
        {claudestyles}
 \bibinfo{year}{[n.\,d.]}\natexlab{}.
\newblock \bibinfo{title}{Understanding {Claude}'s {Personalization} {Features} {\textbar} {Anthropic} {Help} {Center}}.
\newblock
\urldef\tempurl%
\url{https://support.anthropic.com/en/articles/10185728-understanding-claude-s-personalization-features}
\showURL{%
\tempurl}


\bibitem[cha(2025)]%
        {chatgpt-custom}
 \bibinfo{year}{2025}\natexlab{}.
\newblock \bibinfo{title}{{ChatGPT} {Custom} {Instructions}}.
\newblock
\urldef\tempurl%
\url{https://help.openai.com/en/articles/8096356-chatgpt-custom-instructions}
\showURL{%
\tempurl}


\bibitem[gem(2025)]%
        {gemini-personalization}
 \bibinfo{year}{2025}\natexlab{}.
\newblock \bibinfo{title}{Get personalization in {Gemini} {Apps} - {Android} - {Gemini} {Apps} {Help}}.
\newblock
\urldef\tempurl%
\url{https://support.google.com/gemini/answer/15637730?hl=en&co=GENIE.Platform%3DAndroid}
\showURL{%
\tempurl}


\bibitem[cla(2025)]%
        {claude-personalization}
 \bibinfo{year}{2025}\natexlab{}.
\newblock \bibinfo{title}{Understanding {Claude}'s {Personalization} {Features} {\textbar} {Claude} {Help} {Center}}.
\newblock
\urldef\tempurl%
\url{https://support.claude.com/en/articles/10185728-understanding-claude-s-personalization-features}
\showURL{%
\tempurl}


\bibitem[Ahmed et~al\mbox{.}(2025)]%
        {ahmed2025specevalevaluatingmodeladherence}
\bibfield{author}{\bibinfo{person}{Ahmed Ahmed}, \bibinfo{person}{Kevin Klyman}, \bibinfo{person}{Yi Zeng}, \bibinfo{person}{Sanmi Koyejo}, {and} \bibinfo{person}{Percy Liang}.} \bibinfo{year}{2025}\natexlab{}.
\newblock \bibinfo{title}{SpecEval: Evaluating Model Adherence to Behavior Specifications}.
\newblock
\showeprint[arxiv]{2509.02464}~[cs.CL]
\urldef\tempurl%
\url{https://arxiv.org/abs/2509.02464}
\showURL{%
\tempurl}


\bibitem[Alvarado(2023)]%
        {alvarado2023ai}
\bibfield{author}{\bibinfo{person}{Ram{\'o}n Alvarado}.} \bibinfo{year}{2023}\natexlab{}.
\newblock \showarticletitle{AI as an epistemic technology}.
\newblock \bibinfo{journal}{\emph{Science and Engineering Ethics}} \bibinfo{volume}{29}, \bibinfo{number}{5} (\bibinfo{year}{2023}), \bibinfo{pages}{32}.
\newblock


\bibitem[Alves et~al\mbox{.}(2025)]%
        {alves2025benchmarking}
\bibfield{author}{\bibinfo{person}{Jean~V Alves}, \bibinfo{person}{Diogo Leit{\~a}o}, \bibinfo{person}{S{\'e}rgio Jesus}, \bibinfo{person}{Marco~OP Sampaio}, \bibinfo{person}{Javier Li{\'e}bana}, \bibinfo{person}{Pedro Saleiro}, \bibinfo{person}{M{\'a}rio~AT Figueiredo}, {and} \bibinfo{person}{Pedro Bizarro}.} \bibinfo{year}{2025}\natexlab{}.
\newblock \showarticletitle{A benchmarking framework and dataset for learning to defer in human-AI decision-making}.
\newblock \bibinfo{journal}{\emph{Scientific data}} \bibinfo{volume}{12}, \bibinfo{number}{1} (\bibinfo{year}{2025}), \bibinfo{pages}{506}.
\newblock


\bibitem[Chiang et~al\mbox{.}(2024)]%
        {chatbot-arena}
\bibfield{author}{\bibinfo{person}{Wei-Lin Chiang}, \bibinfo{person}{Lianmin Zheng}, \bibinfo{person}{Ying Sheng}, \bibinfo{person}{Anastasios~N. Angelopoulos}, \bibinfo{person}{Tianle Li}, \bibinfo{person}{Dacheng Li}, \bibinfo{person}{Banghua Zhu}, \bibinfo{person}{Hao Zhang}, \bibinfo{person}{Michael~I. Jordan}, \bibinfo{person}{Joseph~E. Gonzalez}, {and} \bibinfo{person}{Ion Stoica}.} \bibinfo{year}{2024}\natexlab{}.
\newblock \showarticletitle{Chatbot arena: an open platform for evaluating LLMs by human preference}. In \bibinfo{booktitle}{\emph{Proceedings of the 41st International Conference on Machine Learning}} (Vienna, Austria) \emph{(\bibinfo{series}{ICML'24})}. \bibinfo{publisher}{JMLR.org}, Article \bibinfo{articleno}{331}, \bibinfo{numpages}{30}~pages.
\newblock


\bibitem[Clark et~al\mbox{.}(2025)]%
        {clark2025epistemic}
\bibfield{author}{\bibinfo{person}{Nicholas Clark}, \bibinfo{person}{Hua Shen}, \bibinfo{person}{Bill Howe}, {and} \bibinfo{person}{Tanu Mitra}.} \bibinfo{year}{2025}\natexlab{}.
\newblock \showarticletitle{Epistemic Alignment: A Mediating Framework for User-{LLM} Knowledge Delivery}. In \bibinfo{booktitle}{\emph{Second Conference on Language Modeling}}.
\newblock
\urldef\tempurl%
\url{https://openreview.net/forum?id=Orvjm9UqH2}
\showURL{%
\tempurl}


\bibitem[Feng et~al\mbox{.}(2024)]%
        {feng-etal-2024-modular}
\bibfield{author}{\bibinfo{person}{Shangbin Feng}, \bibinfo{person}{Taylor Sorensen}, \bibinfo{person}{Yuhan Liu}, \bibinfo{person}{Jillian Fisher}, \bibinfo{person}{Chan~Young Park}, \bibinfo{person}{Yejin Choi}, {and} \bibinfo{person}{Yulia Tsvetkov}.} \bibinfo{year}{2024}\natexlab{}.
\newblock \showarticletitle{Modular Pluralism: Pluralistic Alignment via Multi-{LLM} Collaboration}. In \bibinfo{booktitle}{\emph{Proceedings of the 2024 Conference on Empirical Methods in Natural Language Processing}}, \bibfield{editor}{\bibinfo{person}{Yaser Al-Onaizan}, \bibinfo{person}{Mohit Bansal}, {and} \bibinfo{person}{Yun-Nung Chen}} (Eds.). \bibinfo{publisher}{Association for Computational Linguistics}, \bibinfo{address}{Miami, Florida, USA}, \bibinfo{pages}{4151--4171}.
\newblock
\href{https://doi.org/10.18653/v1/2024.emnlp-main.240}{doi:\nolinkurl{10.18653/v1/2024.emnlp-main.240}}


\bibitem[Fronsdal et~al\mbox{.}(2025)]%
        {petri2025}
\bibfield{author}{\bibinfo{person}{Kai Fronsdal}, \bibinfo{person}{Isha Gupta}, \bibinfo{person}{Abhay Sheshadri}, \bibinfo{person}{Jonathan Michala}, \bibinfo{person}{Stephen McAleer}, \bibinfo{person}{Rowan Wang}, \bibinfo{person}{Sara Price}, {and} \bibinfo{person}{Sam Bowman}.} \bibinfo{year}{2025}\natexlab{}.
\newblock \bibinfo{title}{Petri: Parallel Exploration of Risky Interactions}.
\newblock
\urldef\tempurl%
\url{https://github.com/safety-research/petri}
\showURL{%
\tempurl}


\bibitem[Gao et~al\mbox{.}(2021)]%
        {gao-collab}
\bibfield{author}{\bibinfo{person}{Ruijiang Gao}, \bibinfo{person}{Maytal Saar-Tsechansky}, \bibinfo{person}{Maria De-Arteaga}, \bibinfo{person}{Ligong Han}, \bibinfo{person}{Min~Kyung Lee}, {and} \bibinfo{person}{Matthew Lease}.} \bibinfo{year}{2021}\natexlab{}.
\newblock \showarticletitle{Human-AI Collaboration with Bandit Feedback}. In \bibinfo{booktitle}{\emph{Proceedings of the Thirtieth International Joint Conference on Artificial Intelligence, {IJCAI-21}}}, \bibfield{editor}{\bibinfo{person}{Zhi-Hua Zhou}} (Ed.). \bibinfo{publisher}{International Joint Conferences on Artificial Intelligence Organization}, \bibinfo{pages}{1722--1728}.
\newblock
\href{https://doi.org/10.24963/ijcai.2021/237}{doi:\nolinkurl{10.24963/ijcai.2021/237}}
\newblock
\shownote{Main Track}.


\bibitem[Gehman et~al\mbox{.}(2020)]%
        {gehman2020realtoxicityprompts}
\bibfield{author}{\bibinfo{person}{Samuel Gehman}, \bibinfo{person}{Suchin Gururangan}, \bibinfo{person}{Maarten Sap}, \bibinfo{person}{Yejin Choi}, {and} \bibinfo{person}{Noah~A Smith}.} \bibinfo{year}{2020}\natexlab{}.
\newblock \showarticletitle{RealToxicityPrompts: Evaluating Neural Toxic Degeneration in Language Models}. In \bibinfo{booktitle}{\emph{Findings of the Association for Computational Linguistics: EMNLP 2020}}. \bibinfo{pages}{3356--3369}.
\newblock


\bibitem[Hofer(2000)]%
        {HOFER2000378}
\bibfield{author}{\bibinfo{person}{Barbara~K. Hofer}.} \bibinfo{year}{2000}\natexlab{}.
\newblock \showarticletitle{Dimensionality and Disciplinary Differences in Personal Epistemology}.
\newblock \bibinfo{journal}{\emph{Contemporary Educational Psychology}} \bibinfo{volume}{25}, \bibinfo{number}{4} (\bibinfo{year}{2000}), \bibinfo{pages}{378--405}.
\newblock
\showISSN{0361-476X}
\href{https://doi.org/10.1006/ceps.1999.1026}{doi:\nolinkurl{10.1006/ceps.1999.1026}}


\bibitem[Huang et~al\mbox{.}(2025)]%
        {huang2025survey}
\bibfield{author}{\bibinfo{person}{Lei Huang}, \bibinfo{person}{Weijiang Yu}, \bibinfo{person}{Weitao Ma}, \bibinfo{person}{Weihong Zhong}, \bibinfo{person}{Zhangyin Feng}, \bibinfo{person}{Haotian Wang}, \bibinfo{person}{Qianglong Chen}, \bibinfo{person}{Weihua Peng}, \bibinfo{person}{Xiaocheng Feng}, \bibinfo{person}{Bing Qin}, {et~al\mbox{.}}} \bibinfo{year}{2025}\natexlab{}.
\newblock \showarticletitle{A survey on hallucination in large language models: Principles, taxonomy, challenges, and open questions}.
\newblock \bibinfo{journal}{\emph{ACM Transactions on Information Systems}} \bibinfo{volume}{43}, \bibinfo{number}{2} (\bibinfo{year}{2025}), \bibinfo{pages}{1--55}.
\newblock


\bibitem[Jang et~al\mbox{.}({[n.\,d.]})]%
        {jangpersonalized}
\bibfield{author}{\bibinfo{person}{Joel Jang}, \bibinfo{person}{Seungone Kim}, \bibinfo{person}{Bill~Yuchen Lin}, \bibinfo{person}{Yizhong Wang}, \bibinfo{person}{Jack Hessel}, \bibinfo{person}{Luke Zettlemoyer}, \bibinfo{person}{Hannaneh Hajishirzi}, \bibinfo{person}{Yejin Choi}, {and} \bibinfo{person}{Prithviraj Ammanabrolu}.} \bibinfo{year}{[n.\,d.]}\natexlab{}.
\newblock \showarticletitle{Personalized Soups: Personalized Large Language Model Alignment via Post-hoc Parameter Merging}. In \bibinfo{booktitle}{\emph{Adaptive Foundation Models: Evolving AI for Personalized and Efficient Learning}}.
\newblock


\bibitem[Lai et~al\mbox{.}(2024)]%
        {lai2024large}
\bibfield{author}{\bibinfo{person}{Jinqi Lai}, \bibinfo{person}{Wensheng Gan}, \bibinfo{person}{Jiayang Wu}, \bibinfo{person}{Zhenlian Qi}, {and} \bibinfo{person}{Philip~S Yu}.} \bibinfo{year}{2024}\natexlab{}.
\newblock \showarticletitle{Large language models in law: A survey}.
\newblock \bibinfo{journal}{\emph{AI Open}}  \bibinfo{volume}{5} (\bibinfo{year}{2024}), \bibinfo{pages}{181--196}.
\newblock


\bibitem[Liang et~al\mbox{.}({[n.\,d.]})]%
        {liangholistic}
\bibfield{author}{\bibinfo{person}{Percy Liang}, \bibinfo{person}{Rishi Bommasani}, \bibinfo{person}{Tony Lee}, \bibinfo{person}{Dimitris Tsipras}, \bibinfo{person}{Dilara Soylu}, \bibinfo{person}{Michihiro Yasunaga}, \bibinfo{person}{Yian Zhang}, \bibinfo{person}{Deepak Narayanan}, \bibinfo{person}{Yuhuai Wu}, \bibinfo{person}{Ananya Kumar}, {et~al\mbox{.}}} \bibinfo{year}{[n.\,d.]}\natexlab{}.
\newblock \showarticletitle{Holistic Evaluation of Language Models}.
\newblock \bibinfo{journal}{\emph{Transactions on Machine Learning Research}} (\bibinfo{year}{[n.\,d.]}).
\newblock


\bibitem[Lin et~al\mbox{.}(2022)]%
        {lin2022truthfulqa}
\bibfield{author}{\bibinfo{person}{Stephanie Lin}, \bibinfo{person}{Jacob Hilton}, {and} \bibinfo{person}{Owain Evans}.} \bibinfo{year}{2022}\natexlab{}.
\newblock \showarticletitle{TruthfulQA: Measuring How Models Mimic Human Falsehoods}. In \bibinfo{booktitle}{\emph{Proceedings of the 60th Annual Meeting of the Association for Computational Linguistics (Volume 1: Long Papers)}}. \bibinfo{pages}{3214--3252}.
\newblock


\bibitem[Liu et~al\mbox{.}(2024)]%
        {liu2024survey}
\bibfield{author}{\bibinfo{person}{Lei Liu}, \bibinfo{person}{Xiaoyan Yang}, \bibinfo{person}{Junchi Lei}, \bibinfo{person}{Xiaoyang Liu}, \bibinfo{person}{Yue Shen}, \bibinfo{person}{Zhiqiang Zhang}, \bibinfo{person}{Peng Wei}, \bibinfo{person}{Jinjie Gu}, \bibinfo{person}{Zhixuan Chu}, \bibinfo{person}{Zhan Qin}, {et~al\mbox{.}}} \bibinfo{year}{2024}\natexlab{}.
\newblock \showarticletitle{A Survey on Medical Large Language Models: Technology, Application, Trustworthiness, and Future Directions}.
\newblock \bibinfo{journal}{\emph{CoRR}} (\bibinfo{year}{2024}).
\newblock


\bibitem[Malmqvist(2025)]%
        {malmqvist2025sycophancy}
\bibfield{author}{\bibinfo{person}{Lars Malmqvist}.} \bibinfo{year}{2025}\natexlab{}.
\newblock \showarticletitle{Sycophancy in large language models: Causes and mitigations}. In \bibinfo{booktitle}{\emph{Intelligent Computing-Proceedings of the Computing Conference}}. Springer, \bibinfo{pages}{61--74}.
\newblock


\bibitem[Ojewale et~al\mbox{.}(2025)]%
        {ojewale2025towards}
\bibfield{author}{\bibinfo{person}{Victor Ojewale}, \bibinfo{person}{Ryan Steed}, \bibinfo{person}{Briana Vecchione}, \bibinfo{person}{Abeba Birhane}, {and} \bibinfo{person}{Inioluwa~Deborah Raji}.} \bibinfo{year}{2025}\natexlab{}.
\newblock \showarticletitle{Towards AI accountability infrastructure: Gaps and opportunities in AI audit tooling}. In \bibinfo{booktitle}{\emph{Proceedings of the 2025 CHI Conference on Human Factors in Computing Systems}}. \bibinfo{pages}{1--29}.
\newblock


\bibitem[Shen et~al\mbox{.}(2025)]%
        {shen2025mindvalueactiongapllms}
\bibfield{author}{\bibinfo{person}{Hua Shen}, \bibinfo{person}{Nicholas Clark}, {and} \bibinfo{person}{Tanushree Mitra}.} \bibinfo{year}{2025}\natexlab{}.
\newblock \bibinfo{title}{Mind the Value-Action Gap: Do LLMs Act in Alignment with Their Values?}
\newblock
\showeprint[arxiv]{2501.15463}~[cs.HC]
\urldef\tempurl%
\url{https://arxiv.org/abs/2501.15463}
\showURL{%
\tempurl}


\bibitem[Sorensen et~al\mbox{.}(2024)]%
        {sorensen2024position}
\bibfield{author}{\bibinfo{person}{Taylor Sorensen}, \bibinfo{person}{Jared Moore}, \bibinfo{person}{Jillian Fisher}, \bibinfo{person}{Mitchell Gordon}, \bibinfo{person}{Niloofar Mireshghallah}, \bibinfo{person}{Christopher~Michael Rytting}, \bibinfo{person}{Andre Ye}, \bibinfo{person}{Liwei Jiang}, \bibinfo{person}{Ximing Lu}, \bibinfo{person}{Nouha Dziri}, {et~al\mbox{.}}} \bibinfo{year}{2024}\natexlab{}.
\newblock \showarticletitle{Position: a roadmap to pluralistic alignment}. In \bibinfo{booktitle}{\emph{Proceedings of the 41st International Conference on Machine Learning}}. \bibinfo{pages}{46280--46302}.
\newblock


\bibitem[wai Chan~* and Elliott(2004)]%
        {Chan}
\bibfield{author}{\bibinfo{person}{Kwok wai Chan~*} {and} \bibinfo{person}{Robert~G. Elliott}.} \bibinfo{year}{2004}\natexlab{}.
\newblock \showarticletitle{Epistemological beliefs across cultures: critique and analysis of beliefs structure studies}.
\newblock \bibinfo{journal}{\emph{Educational Psychology}} \bibinfo{volume}{24}, \bibinfo{number}{2} (\bibinfo{year}{2004}), \bibinfo{pages}{123--142}.
\newblock
\showeprint{https://doi.org/10.1080/0144341032000160100}
\href{https://doi.org/10.1080/0144341032000160100}{doi:\nolinkurl{10.1080/0144341032000160100}}


\bibitem[Wang et~al\mbox{.}(2024)]%
        {wang2024large}
\bibfield{author}{\bibinfo{person}{Shen Wang}, \bibinfo{person}{Tianlong Xu}, \bibinfo{person}{Hang Li}, \bibinfo{person}{Chaoli Zhang}, \bibinfo{person}{Joleen Liang}, \bibinfo{person}{Jiliang Tang}, \bibinfo{person}{Philip~S Yu}, {and} \bibinfo{person}{Qingsong Wen}.} \bibinfo{year}{2024}\natexlab{}.
\newblock \showarticletitle{Large language models for education: A survey and outlook}.
\newblock \bibinfo{journal}{\emph{arXiv preprint arXiv:2403.18105}} (\bibinfo{year}{2024}).
\newblock


\bibitem[Zhang et~al\mbox{.}(2025)]%
        {zhang2025personalization}
\bibfield{author}{\bibinfo{person}{Zhehao Zhang}, \bibinfo{person}{Ryan~A. Rossi}, \bibinfo{person}{Branislav Kveton}, \bibinfo{person}{Yijia Shao}, \bibinfo{person}{Diyi Yang}, \bibinfo{person}{Hamed Zamani}, \bibinfo{person}{Franck Dernoncourt}, \bibinfo{person}{Joe Barrow}, \bibinfo{person}{Tong Yu}, \bibinfo{person}{Sungchul Kim}, \bibinfo{person}{Ruiyi Zhang}, \bibinfo{person}{Jiuxiang Gu}, \bibinfo{person}{Tyler Derr}, \bibinfo{person}{Hongjie Chen}, \bibinfo{person}{Junda Wu}, \bibinfo{person}{Xiang Chen}, \bibinfo{person}{Zichao Wang}, \bibinfo{person}{Subrata Mitra}, \bibinfo{person}{Nedim Lipka}, \bibinfo{person}{Nesreen~K. Ahmed}, {and} \bibinfo{person}{Yu Wang}.} \bibinfo{year}{2025}\natexlab{}.
\newblock \showarticletitle{Personalization of Large Language Models: A Survey}.
\newblock \bibinfo{journal}{\emph{Transactions on Machine Learning Research}} (\bibinfo{year}{2025}).
\newblock
\showISSN{2835-8856}
\urldef\tempurl%
\url{https://openreview.net/forum?id=tf6A9EYMo6}
\showURL{%
\tempurl}
\newblock
\shownote{Survey Certification}.


\end{thebibliography}


\end{document}